\documentclass[aps,prc,twocolumn,superscriptaddress]{revtex4-1}

\usepackage{graphicx}
\usepackage{booktabs}

\begin{document}

\title{Mixed-symmetry octupole and hexadecapole excitations in the $N=52$ isotones}

\author{A.~Hennig}
\email[]{hennig@ikp.uni-koeln.de}
%\homepage[]{Your web page}
%\thanks{}
%\altaffiliation{}
\affiliation{Institut f\"ur Kernphysik, Universit\"at zu K\"oln, D-50937 K\"oln, Germany}

\author{M.~Spieker}
\affiliation{Institut f\"ur Kernphysik, Universit\"at zu K\"oln, D-50937 K\"oln, Germany}

\author{V.~Werner}
\affiliation{Wright Nuclear Structure Laboratory, Yale University, New Haven, Connecticut 06520, USA}
\affiliation{Institut f\"ur Kernphysik, Technische Universit\"at Darmstadt, D-6489 Darmstadt, Germany}

\author{T.~Ahn}
\altaffiliation[Present address: ]{Department of Physics, University of Notre Dame, Notre Dame, Indiana 46556, USA.}
\affiliation{Wright Nuclear Structure Laboratory, Yale University, New Haven, Connecticut 06520, USA}

\author{V.~Anagnostatou}
\affiliation{Wright Nuclear Structure Laboratory, Yale University, New Haven, Connecticut 06520, USA}
\affiliation{Department of Physics, University of Surrey, Guildford, GU2 7XH, UK}

\author{N.~Cooper}
\affiliation{Wright Nuclear Structure Laboratory, Yale University, New Haven, Connecticut 06520, USA}

\author{V.~Derya}
\affiliation{Institut f\"ur Kernphysik, Universit\"at zu K\"oln, D-50937 K\"oln, Germany}

\author{M.~Elvers}
\affiliation{Institut f\"ur Kernphysik, Universit\"at zu K\"oln, D-50937 K\"oln, Germany}
\affiliation{Wright Nuclear Structure Laboratory, Yale University, New Haven, Connecticut 06520, USA}

\author{J.~Endres}
\affiliation{Institut f\"ur Kernphysik, Universit\"at zu K\"oln, D-50937 K\"oln, Germany}

\author{P.~Goddard}
\affiliation{Wright Nuclear Structure Laboratory, Yale University, New Haven, Connecticut 06520, USA}
\affiliation{Department of Physics, University of Surrey, Guildford, GU2 7XH, UK}

\author{A.~Heinz}
\affiliation{Wright Nuclear Structure Laboratory, Yale University, New Haven, Connecticut 06520, USA}
\affiliation{Fundamental Fysik, Chalmers Tekniska H\"ogskola, SE-41296 G\"oteborg, Sweden}

\author{R.~O.~Hughes}
\affiliation{Wright Nuclear Structure Laboratory, Yale University, New Haven, Connecticut 06520, USA}
\affiliation{University of Richmond, Richmond, Virginia 23173, USA}

\author{G.~Ilie}
\affiliation{Wright Nuclear Structure Laboratory, Yale University, New Haven, Connecticut 06520, USA}
\affiliation{National Institute for Physics and Nuclear Engineering, Bucharest-Magurele, RO-77125, Romania}

\author{M.~N.~Mineva}
\affiliation{Institute for Nuclear Research and Nuclear Energy, Bulgarian Academy of Sciences, BG-1784 Sofia, Bulgaria}

\author{P.~Petkov}
\affiliation{Institut f\"ur Kernphysik, Universit\"at zu K\"oln, D-50937 K\"oln, Germany}
\affiliation{Institute for Nuclear Research and Nuclear Energy, Bulgarian Academy of Sciences, BG-1784 Sofia, Bulgaria}

\author{S.~G.~Pickstone}
\affiliation{Institut f\"ur Kernphysik, Universit\"at zu K\"oln, D-50937 K\"oln, Germany}

\author{N.~Pietralla}
\affiliation{Institut f\"ur Kernphysik, Technische Universit\"at Darmstadt, D-6489 Darmstadt, Germany}

\author{D.~Radeck}
\affiliation{Institut f\"ur Kernphysik, Universit\"at zu K\"oln, D-50937 K\"oln, Germany}
\affiliation{Wright Nuclear Structure Laboratory, Yale University, New Haven, Connecticut 06520, USA}

\author{T.~J.~Ross}
\affiliation{Department of Physics, University of Surrey, Guildford, GU2 7XH, UK}
\affiliation{University of Richmond, Richmond, Virginia 23173, USA}

\author{D.~Savran}
\affiliation{ExtreMe Matter Institute EMMI and Research Division, GSI, D-64291 Darmstadt, Germany}
\affiliation{Frankfurt Institute for Advanced Studies FIAS, D-60438 Frankfurt a.M., Germany}

\author{A.~Zilges}
\affiliation{Institut f\"ur Kernphysik, Universit\"at zu K\"oln, D-50937 K\"oln, Germany}

\date{\today}

\begin{abstract}

\noindent\textbf{Background:} Excitations with mixed proton-neutron symmetry have been previously observed in the $N=52$ isotones. Besides the well established quadrupole mixed-symmetry states (MSS), octupole and hexadecapole MSS have been recently proposed for the nuclei $^{92}$Zr and $^{94}$Mo.

\noindent\textbf{Purpose:} The heaviest stable $N=52$ isotone $^{96}$Ru was investigated to study the evolution of octupole and hexadecapole MSS with increasing proton number.

\noindent\textbf{Methods:} Two inelastic proton-scattering experiments on $^{96}$Ru were performed to extract branching ratios, multipole mixing ratios, and level lifetimes. From the combined data, absolute transition strengths were calculated.

\noindent\textbf{Results:} Strong $M1$ transitions between the lowest-lying $3^-$ and $4^+$ states were observed, providing evidence for a one-phonon mixed-symmetry character of the $3^{(-)}_2$ and $4^+_2$ states.

\noindent\textbf{Conclusions:} $sdg$-IBM-2 calculations were performed for $^{96}$Ru. The results are in excellent agreement with the experimental data, pointing out a one-phonon hexadecapole mixed-symmetry character of the $4^+_2$ state. The $\big< 3^-_1\left||M1\right||3^{(-)}_2\big>$ matrix element is found to scale with the $\left<2^+_{\mathrm{s}}\left||M1\right||2^+_{\mathrm{ms}}\right>$ matrix element.

\end{abstract}

% insert suggested PACS numbers in braces on next line
% insert suggested keywords - APS authors don't need to do this

\maketitle

\textit{Introduction.} Protons and neutrons are the building blocks of atomic nuclei, which feature collective excitations which are symmetric or not symmetric with respect to the proton-neutron degree of freedom \cite{Heyd86}. Excitations resulting from the antisymmetric coupling of proton and neutron eigenstates are usually referred to as mixed-symmetry states (MSS), whereas the symmetric coupling results in fully-symmetric states (FSS) \cite{Iach84}. Mixed-symmetry quadrupole excitations are predicted within the $sd$ proton-neutron version of the interacting boson model ($sd$-IBM-2) \cite{Arim75, Arim77, Otsu78, Isac86}, where $s$- and $d$-bosons are obtained by coupling protons and neutrons to pairs with angular momentum $L=0$ and $L=2$, respectively. In the IBM-2, MSS and FSS can be distinguished by their $F$-spin quantum number \cite{Arim77, Otsu78}, which is the bosonic analog of isospin for fermions. Strong $F$-vector ($\Delta F=1$) $M1$ transitions from MSS to their symmetric counterparts are predicted by the model. In the IBM-1, where proton and neutron bosons are not distinguished, $M1$ transitions with a one-body $M1$ transition operator are forbidden. Thus, $M1$ transitions serve as a key signature for MSS \cite{Heyd10, Piet08}.

Mixed-symmetry quadrupole excitations are well established in the stable $N=52$ isotones \cite{Piet99b, Piet00, Piet01, Wern02, Klei02, Fran03, Fran05, Linn05a, Orce06, Piet08}, see Ref.~\cite{Piet08} for a review. In addition, the existence of higher-order multipolarity mixed-symmetry states has been recently proposed for the $N=52$ isotones $^{92}$Zr and $^{94}$Mo, namely, of octupole ($L=3$) \cite{Fran03, Smir00, Sche10} and hexadecapole ($L=4$) character \cite{Fran03, Fran05, Casp13}. As for the quadrupole MSS, experimental evidence came from the observation of remarkably strong $M1$ transition strengths in the order of $\sim 1~\mu_N^2$ between the lowest-lying $3^-$ and $4^+$ states, respectively. 

Candidates for octupole excitations with mixed-symmetry character have been proposed in various nuclei in the $A\approx 100$ mass region \cite{Sche10}, among others also in the $N=52$ isotones $^{92}$Zr and $^{94}$Mo \cite{Fran05, Fran03}. MS octupole excitations were predicted in $sdf$-IBM-2 calculations \cite{Smir00}. Along with the $M1$ fingerprint, a sizable $E1$ transition to the FS one-phonon quadrupole state $2^+_{\mathrm{s}}$ is expected in the $U_{\pi\nu}(1)\otimes U_{\pi\nu}(5)\otimes U_{\pi\nu}(7)$ limit, according to the two-body nature of the $E1$ operator \cite{Piet03, Sche10}. In addition, a strong $E1$ transition to the MS one-phonon quadrupole state $2^+_{\mathrm{ms}}$ has been observed in the case of $^{94}$Mo.

Recently, the strong $M1$ transition between the lowest-lying $4^+$ states of $^{94}$Mo was successfully reproduced within the $sdg$-IBM-2 without abandoning the description of quadrupole MSS \cite{Casp13}, suggesting the strong $M1$ transition to result from MS and FS one-phonon hexadecapole components in the $4^+_2$ and $4^+_1$ states, respectively. Additional evidence for this interpretation is provided by shell-model calculations for $^{92}$Zr and $^{94}$Mo \cite{Wern02, Lise00, Fran03}, indicating dominant $\nu=2$, $j=4$ configurations for the lowest-lying $4^+$ states. These are by definition identified with $g$-bosons in the IBM; here, $\nu$ denotes the seniority.

The study in Ref.~\cite{Casp13} was based on the only experimentally known case at that time. The intention of the present work is to show, that  the case of $^{94}$Mo is not exceptional, but that the presence of hexadecapole components in the wave functions of low-lying $4^+$ states in near-spherical nuclei is a general phenomenon. For this purpose, we have studied the heaviest stable $N=52$ isotone $^{96}$Ru in two proton-scattering experiments. In addition, the structure of the low-lying $4^+$ states was investigated in the framework of $sdg$-IBM-2 calculations. Details on the experimental aspects will be given in a more extensive article.

\textit{Experiments.} To identify MSS based on absolute transition strength, two inelastic proton scattering experiments were performed. The first one at the Wright Nuclear Structure Laboratory (WNSL) at Yale University, USA, the second one at the Institute for Nuclear Physics at the University of Cologne, Germany. 

In the former, an $8.4~\mathrm{MeV}$ proton beam, provided by the ESTU Tandem Accelerator, impinged on a $106~\mathrm{\mu g/cm^2}$ enriched $^{96}$Ru target, supported by a $^{12}$C backing with a thickness of $14~\mathrm{\mu g/cm^2}$. The scattered protons were detected using five silicon surface-barrier detectors, positioned predominantly at backward angles. For the $\gamma$-ray detection, eight BGO-shielded Clover-type HPGe detectors of the YRAST Ball spectrometer \cite{Beau00} were used. Further information on the experimental setup can be found in Ref.~\cite{Elve11}. From the energy information of scattered protons, the excitation energy $E_x$ was deduced on an event-by-event basis. Thus, $\gamma$-decay branching ratios were extracted with high sensitivity from the acquired $p\gamma$ coincidence data by gating on a specific excitation energy \cite{Wilh96}. Spin quantum numbers and multipole mixing ratios were obtained by means of the $\gamma\gamma$ angular-correlation technique \cite{Kran73}.

For the extraction of level lifetimes, a second proton-scattering experiment was performed at the Institute for Nuclear Physics at the University of Cologne. The same target was bombarded with a $7.0~\mathrm{MeV}$ proton beam, provided by the 10~MV FN Tandem accelerator. For the coincident detection of the scattered protons and de-exciting $\gamma$-rays, the particle-detector array SONIC, equipped with six passivated implanted planar silicon (PIPS) detectors, was embedded within the $\gamma$-ray spectrometer HORUS. Nuclear level lifetimes were measured by means of the Doppler-shift attenuation method (DSAM) \cite{Alex78, Petk98} using $p\gamma$ coincidence data \cite{Seam69}. Peak centroids were extracted from $\gamma$-ray spectra that were gated on the excitation energy of the level of interest. This way, feeding from higher-lying states is eliminated. The stopping process of the recoil nuclei in the target and stopper material was modeled by means of a Monte-Carlo simulation \cite{Curr69} using the computer code \textsc{dstop96} \cite{Petk98} which is based on the code \textsc{desastop} \cite{Wint83a}. More detailed information on the experimental technique and the data analysis will be the subject of an upcoming publication. Absolute transition strengths were finally calculated from the combined experimental data of both experiments.

\textit{Mixed-symmetry octupole excitations.} For the $J=3$ state of $^{96}$Ru at $3076~\mathrm{keV}$, a negative parity has been previously assigned based on the observation of a $\gamma$ decay to the $5^-$ state at $2588~\mathrm{keV}$ \cite{Adam86}. As in \cite{Klei02}, this $\gamma$ decay was not confirmed in the present experiments. However, since the $3^-_{\mathrm{ms}}$ candidates of $^{94}$Mo ($3011~\mathrm{keV}$) and $^{92}$Zr ($3040~\mathrm{keV}$) have been observed at similar excitation energies, a negative parity was assigned tentatively. With this assumption, an $M1$ transition strength of $B(M1)=0.14(4)~\mathrm{\mu_N^2}$ was obtained for the $3^{(-)}_2\rightarrow~3^-_1$ transition. Therewith, the $3^{(-)}_2$ state is a likely candidate for the one-phonon MS octupole state. As for the case of $^{94}$Mo, an $E1$ strength of $B(E1)=0.14(3)~\mathrm{mW.u.}$ to the known $2^+_{\mathrm{ms}}$ state at $E_x=2283~\mathrm{keV}$ was obtained. However, only a weak $E1$ strength of $B(E1)=0.0017(3)~\mathrm{mW.u.}$ was extracted for the $3^{(-)}_2\rightarrow~2^+_{\mathrm{s}}$ transition.

%In the $U_{\pi\nu}(1)\otimes U_{\pi\nu}(5)\otimes U_{\pi\nu}(7)$ limit of the $sdf$-IBM-2, the reduced transition strength of the $3^-_{\mathrm{ms}}\rightarrow 3^-_1$ transition can be calculated analytically \cite{Smir00}:
%\begin{equation}
%B(M1;3^-_{\mathrm{ms}}\rightarrow 3^-_1) = \frac{9}{\pi}\left(g_{\pi}-g_{\nu}\right)^2\frac{N_{\pi}N_{\nu}}{N^2}
%\end{equation}
%with $g_{\pi}$ and $g_{\nu}$ being the proton and neutron $g$-factors for the $f$-bosons and $N_{\pi}$, $N_{\nu}$, and $N$, the proton, neutron, and total boson numbers, respectively. Assuming $^{100}$Sn as inert core gives $N_{\pi}=3$ and $N_{\nu}=1$. With the bare orbital $g$-factors for protons and neutrons ($g_{\pi}=1$ and $g_{\nu}=0$), a reduced $M1$ transition strength of $0.54~\mu_N^2$ is predicted, which is almost a factor of four larger compared to the experimental value.

As expected for collective excitations, the $3^-_2\rightarrow~3^-_1$ $M1$ matrix element scales with the one for the $2^+_{\mathrm{ms}}\rightarrow~2^+_{\mathrm{s}}$ transition for several nuclei in the $A\approx 100$ mass region \cite{Sche10}, in particular also for the $N=52$ isotones $^{92}$Zr and $^{94}$Mo. With the bare $g$ factors ($g_{\pi}=1$ and $g_{\nu}=0$), a value of $\sqrt{14/5}\approx 1.67$ is predicted in the $U_{\pi\nu}(1)\otimes U_{\pi\nu}(5)\otimes U_{\pi\nu}(7)$ limit of the $sdf$-IBM-2 for the ratio of the matrix elements \cite{Smir00}. The experimental ratios are close to unity but stay rather constant \cite{Sche10}. Only the value for $^{96}$Mo deviates from the others by a factor of 2. From our new data, we calculated a ratio of $\frac{\left<3^-_1\left||M1\right||3^{(-)}_2\right>}{\left<2^+_{\mathrm{s}}\left||M1\right||2^+_{\mathrm{ms}}\right>}=0.53(9)$ for $^{96}$Ru, close to the value for $^{96}$Mo. The deviation of the ratio for $^{96}$Ru compared to the values for the other $N=52$ isotones might result from the more $O(6)$-like structure of $^{96}$Ru compared to, e.g., $^{94}$Mo (see below).

\begin{figure}[t!]
	\includegraphics[width=0.5\textwidth]{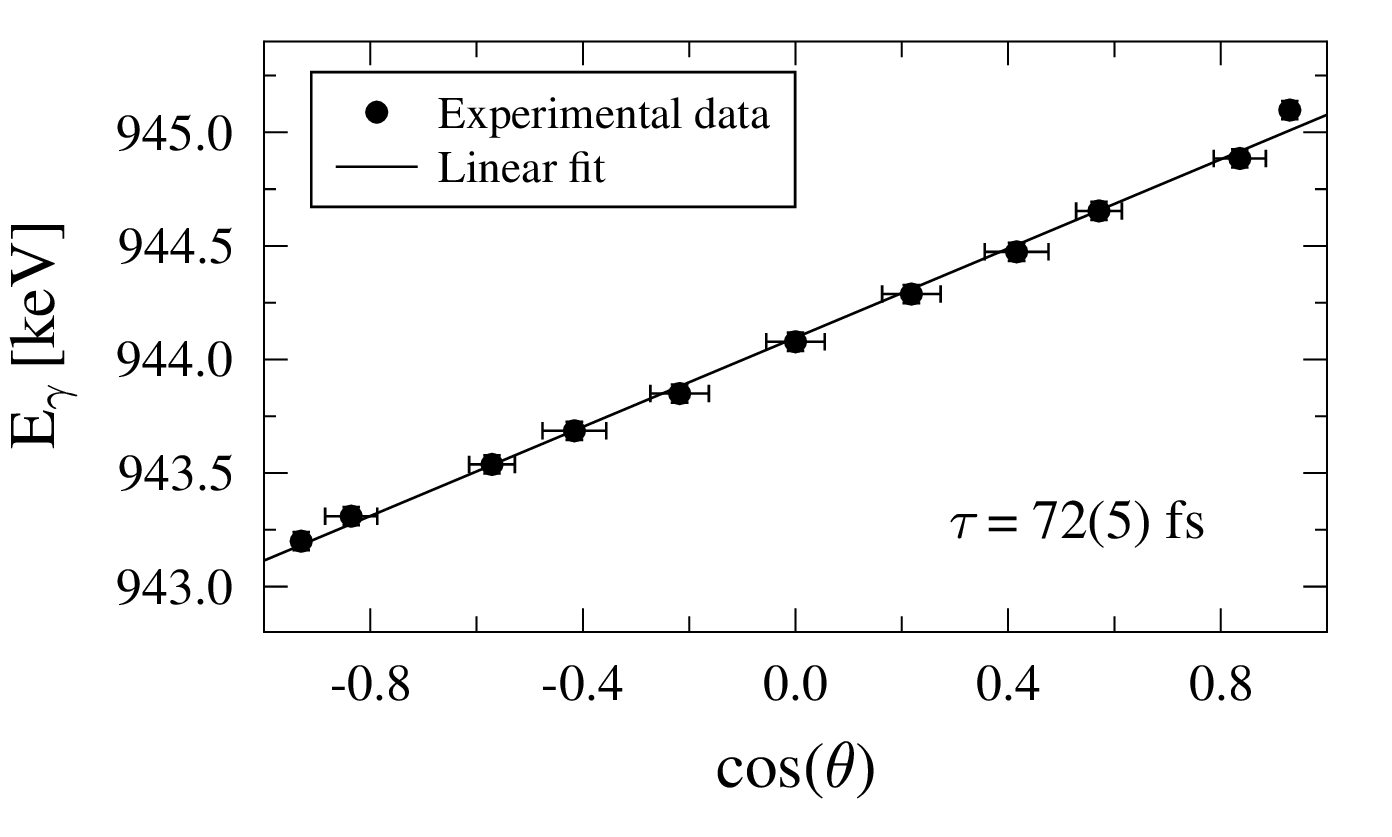}
\caption{\label{fig:shift}Centroid shift of the $E_{\gamma}^0=944~\mathrm{keV}$ $4^+_2\rightarrow~4^+_1$ $\gamma$-transition of $^{96}$Ru as a function of $\cos(\theta)$. $\theta$ is the angle between the recoil direction of motion and the direction in which the $\gamma$ ray is emitted. From the slope, the Doppler-shift attenuation factor is calculated.}
\end{figure}

\textit{Hexadecapole excitations.} For the $J=4$ state of $^{96}$Ru at $E_x=2462~\mathrm{keV}$, a positive parity was assigned because of a newly observed $\gamma$-decay to the $2^+_1$ state. A lifetime of $\tau=140^{+70}_{-40}~\mathrm{fs}$ has been previously reported for this state \cite{Adam86}, characterized by large uncertainties in the determination of the Doppler-shift attenuation factor. From our present analysis a lifetime of $\tau=72(5)~\mathrm{fs}$ was extracted. Figure~\ref{fig:shift} shows the centroid energy of the $E_{\gamma}^0=944~\mathrm{keV}$ $4^+_2\rightarrow~4^+_1$ $\gamma$-transition as a function of $\cos(\theta)$, where $\theta$ is the angle between the initial direction of motion of the recoil nucleus and the direction of the $\gamma$-ray emission. For the transition to the $4^+_1$ state, an $M1$ transition strength of $0.90(18)~\mathrm{\mu_N^2}$ was derived, which is even stronger than the $M1$ strength of the $2^+_{\mathrm{ms}}\rightarrow~2^+_{\mathrm{s}}$ transition \cite{Piet01}. The $M1$ strength between the lowest-lying $4^+$ states of $^{94}$Mo is of comparable size \cite{Fran03}. Hence, the $4^+_2$ state is a likely candidate to show one-phonon hexadecapole MS contributions.

\textit{$sdg$-IBM-2 calculations.} The first $sdg$-IBM-2 calculations on the $N=52$ isotones were performed by Casperson \textit{et al.} for the nucleus $^{94}$Mo \cite{Casp13}. For the first time, the strong $M1$ transition between the lowest-lying $4^+$ states in $^{94}$Mo was reproduced without deteriorating the description of the well established quadrupole mixed-symmetry features. Motivated by this work, we chose the same Hamiltonian and transition operators for the description of $^{96}$Ru:

\begin{eqnarray}
\hat{H}&=&c\Big\{(1-\zeta)\big(\hat{n}_{d_{\pi}}+\hat{n}_{d_{\nu}}+\alpha (\hat{n}_{g_{\pi}}+\hat{n}_{g_{\nu}})\big)\nonumber \\
 & & -\frac{\zeta}{4N}(\hat{Q}_{\pi}+\hat{Q}_{\nu})\cdot (\hat{Q}_{\pi}+\hat{Q}_{\nu})\nonumber \\
 & & +\lambda_{sd}\hat{M}_{sd}+\lambda_{sg}\hat{M}_{sg}\Big\},
\end{eqnarray}

with 
\begin{eqnarray}
\hat{Q}_{\rho} &=& [s_{\rho}^{\dagger}\tilde{d}_{\rho}+d_{\rho}^{\dagger}\tilde{s}_{\rho}]^{(2)} +\beta[d_{\rho}^{\dagger}\tilde{g}_{\rho}+g_{\rho}^{\dagger}\tilde{d}_{\rho}]^{(2)}+\nonumber \\
& & \chi_{d}[d_{\rho}^{\dagger}\tilde{d}_{\rho}]^{(2)}+\chi_{g}[g_{\rho}^{\dagger}\tilde{g}_{\rho}]^{(2)},
\end{eqnarray}

and $\rho=\pi,\nu$. The $M1$ and $E2$ transition operators are defined as

\begin{equation}
\hat{T}(M1)=\sqrt{\frac{3}{4\pi}}\left(g_{d_{\pi}}\hat{L}_{d_{\pi}}+g_{d_{\nu}}\hat{L}_{d_{\nu}}+g_{g_{\pi}}\hat{L}_{g_{\pi}}+g_{g_{\nu}}\hat{L}_{g_{\nu}}\right)
\end{equation}
and 
\begin{equation}
\hat{T}(E2)=e_{B_{\pi}}\hat{Q}_{\pi}+e_{B_{\nu}}\hat{Q}_{\nu}~,
\end{equation}

respectively. For detailed information on the Hamiltonian and the transition operators, see \cite{Casp13}. The calculations were performed with the computer code \textsc{ArbModel} \cite{Hein08}. The number of valence bosons was chosen with respect to $^{100}$Sn as inert core, resulting in $N_{\pi}=3$ and $N_{\nu}=1$. 

\begin{table}[t]
\caption{\label{tab:params}$sdg$-IBM-2 parameters obtained from the parameter scan for $^{96}$Ru compared with the parameters obtained for $^{94}$Mo \cite{Casp13}. $e_{B_{\pi}}$ is quoted in units of $~\sqrt{\mathrm{W.u.}}$, $g_{\rho_{\pi}}$ is quoted in units of $\mu_N$. See also \cite{Casp13} for details.}
\begin{ruledtabular}
\begin{tabular}{ccc}
\textrm{Parameter}&
\textrm{$^{96}$Ru}&
\textrm{$^{94}$Mo}\\
\colrule
$c$							& $5.58$		& $3.53$\\		
$\zeta$						& $0.78$		& $0.64$\\
$\alpha$					& $1.1$			& $1.4$\\
$\beta$						& $1.5$			& $1.86$\\
$\lambda_{sd}$				& $0.026$		& $0.05$\\
$\lambda_{sg}$				& $0.018$		& $0.016$\\
$e_{B_{\pi}}$ 				& $2.26$		& $1.83$\\
$g_{\pi}$		 			& $1.34$		& $1.44$\\

\end{tabular}
\end{ruledtabular}
\end{table}

To reduce the number of parameters, the proton $g$-factors $g_{d_{\pi}}$ and  $g_{g_{\pi}}$ were set equal and the neutron effective charges $e_{B_{\nu}}$ and $g$ factors $g_{d_{\nu}}$ and $g_{g_{\nu}}$ were set to zero, as were the parameters $\chi_d$ and $\chi_g$. To fix the remaining five free parameters of the Hamiltonian, a parameter scan was performed to optimize the calculation to reproduce the energy of the $2^+_1$ state, the $R_{4/2}$ ratio, the $B(E2;4^+_1\rightarrow~2^+_1)/B(E2;2^+_1\rightarrow~0^+_1)$ ratio, the energy of the known one-phonon quadrupole MSS, which is the $2^+_3$ state of $^{96}$Ru \cite{Piet01}, and the $B(M1;4^+_2\rightarrow~4^+_1)/B(M1;2^+_3\rightarrow~2^+_1)$ ratio. 

The effective charges $e_{B_{\rho}}$ set the scale for $E2$ transitions and were fixed to reproduce the $B(E2;2^+_1\rightarrow~0^+_1)$ value. The $g$ factor $g_{\pi}=g_{d_{\pi}}=g_{g_{\pi}}$  was fixed to describe the $B(M1;2^+_3\rightarrow~2^+_1)$ value. The obtained parameters are quoted in Table \ref{tab:params} and are similar to those obtained for $^{94}$Mo \cite{Casp13}. The larger value of $\zeta=0.78$ for $^{96}$Ru indicates a more $O(6)$-like structure compared to $^{94}$Mo. Only little difference is found for the excitation energies of the $d$ and $g$ bosons for $^{96}$Ru, governed by the parameter $\alpha$.

The $sdg$-IBM-2 with the chosen Hamiltonian and parameters is a simplified approach. A more sophisticated description has to include for example non-vanishing parameters $\chi_{d,g}^{\rho}$ to allow $SU(3)$ contributions. However, the choice of $\chi^{\pi}_{d,g}\neq\chi^{\nu}_{d,g}$ would result in a breaking of $F$-spin symmetry, which was avoided to maintain a clear distinction between MSS and FSS. In addition, the chosen Hamiltonian conserves $d$-parity \cite{Piet98}.

\begin{figure}[t!]
	\includegraphics[width=0.45\textwidth]{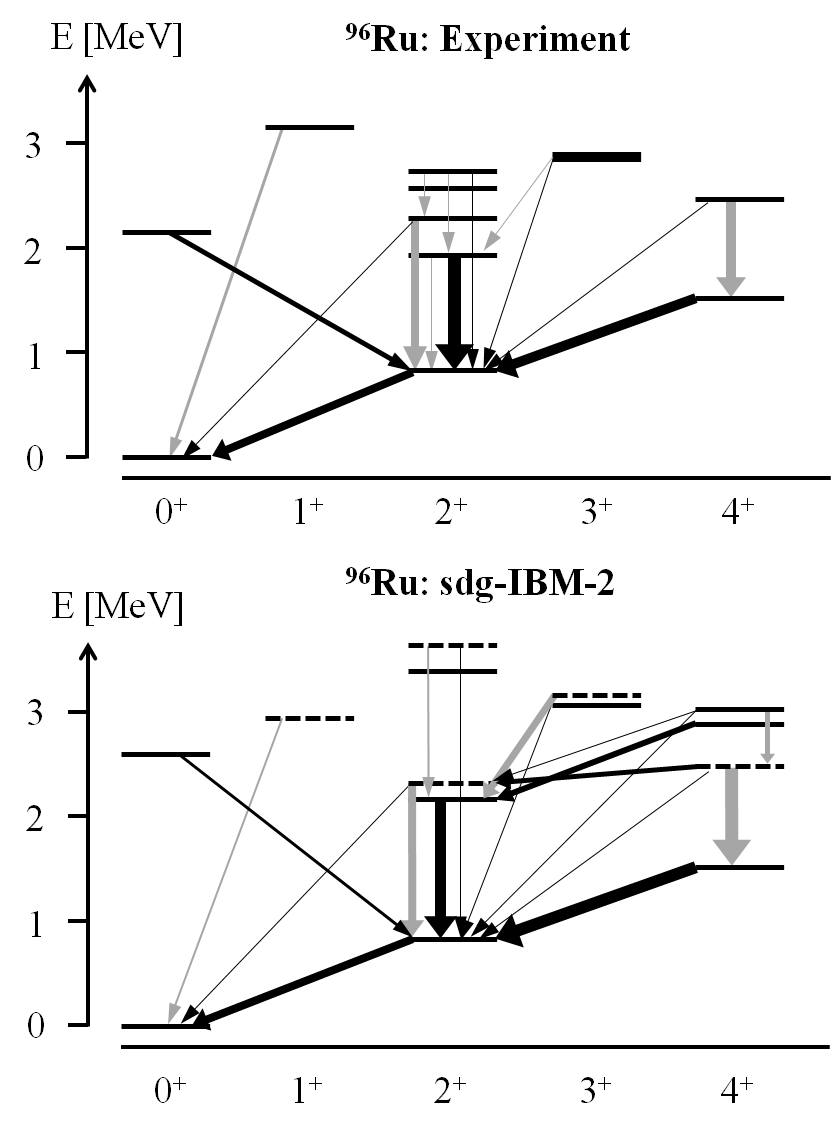}
\caption{\label{fig:levels}Comparison of experimental (upper panel) and calculated (lower panel) level schemes for positive-parity low-spin states of $^{96}$Ru. $M1$ and $E2$ transitions are indicated by gray and black arrows, respectively. The widths of the arrows are proportional to the transition strengths. States for which the IBM predicts an $F$-spin quantum number of $F_{\mathrm{max}}-1$ are marked with dashed lines.}
\end{figure}

The calculated level scheme is in good agreement with the data, as shown in Fig. \ref{fig:levels}. In particular, the excitation energies of the $4^+_{1,2}$ states are well reproduced. Only for the excitation energies of the $2^+_4$ and $2^+_5$ states significant deviations from the experimental values were obtained.

\begin{table}[t]
\caption{\label{tab:trans}Experimental level energies and $E2$ and $M1$ transition strengths of $^{96}$Ru compared with the results from $sdg$-IBM-2 calculations. The predicted $E4$ transition strengths for the decay of the lowest $4^+$ states to the ground state are shown as well. $E2$ and $E4$ strengths are given in $\mathrm{W.u.}$, $M1$ transitions are quoted in units of $\mu_N^2$. If not indicated differently, the experimental values were obtained in this work. The $F$-spin quantum number predicted by the IBM is shown in the second column.}
\begin{ruledtabular}
\begin{tabular}{ccccclll}
&	&	\multicolumn{2}{c}{Energies} &	\multicolumn{4}{c}{Transition strengths B($\sigma\lambda$)}	\\
&	$F$ & $E_{\mathrm{Exp}}$	& $E_{\mathrm{IBM}}$	&	$J^{\pi}_i \rightarrow J^{\pi}_f$	&	$\sigma\lambda$	&	Exp.	&	IBM \\
\colrule
$0^+_1$	&	2	&	0.000	&	0.000	&	-	&	-	&	-	&	-	\\
%\addlinespace[0.2cm]
$1^+_1$	&	1	&	3.154	&	2.944	&	$1^+_1\rightarrow 0^+_1$	&	$M1$	&	0.17(5)\footnote{A value of $0.30(4)~\mathrm{\mu_N^2}$ was reported in \cite{Linn05a}.}		&	0.13	\\
%\addlinespace[0.2cm]
$2^+_1$	&	2	&	0.832	&	0.832	&	$2^+_1\rightarrow 0^+_1$	&	$E2$	&	18.1(5)\footnote{Adopted from \cite{Klei02}.}		&	18.4	\\
%\addlinespace[0.2cm]
$2^+_2$	&	2	&	1.932	&	2.165	&	$2^+_2\rightarrow 2^+_1$	&	$M1$	&	0.05(2)		&	0		\\	
		&		&			&			&	$2^+_2\rightarrow 2^+_1$	&	$E2$	&	28(9)\footnote{A value of $19(4)~\mathrm{W.u.}$ was reported in \cite{Klei02}.}			&	24		\\
%\addlinespace[0.2cm]
$2^+_3$	&	1	&	2.283	&	2.322	&	$2^+_3\rightarrow 2^+_1$	&	$M1$	&	0.69(14)\footnote{A value of $0.78(23)~\mathrm{\mu_N^2}$ was reported in \cite{Piet01}.}	&	0.69	\\
		&		&			&			&	$2^+_3\rightarrow 0^+_1$	&	$E2$	&	1.36(19)	&	2.53	\\
%\addlinespace[0.2cm]
$3^+_1$	&	2	&	2.852	&	3.072	&	$3^+_1\rightarrow 2^+_1$	&	$E2$	&	$<0.01$		&	0		\\
		&		&			&			&	$3^+_1\rightarrow 2^+_1$	&	$M1$	&	0.008(1)	&	0		\\
		&		&			&			&	$3^+_1\rightarrow 2^+_2$	&	$E2$	&	$<5.58$		&	14.7	\\
%\addlinespace[0.2cm]
$3^+_2$	&	1	&	2.898	&	3.158	&	$3^+_2\rightarrow 2^+_1$	&	$E2$	&	$<0.28$		&	3.17	\\
		&		&			&			&	$3^+_2\rightarrow 2^+_2$	&	$E2$	&	0.02(4)		&	0		\\
		&		&			&			&	$3^+_2\rightarrow 2^+_2$	&	$M1$	&	0.078(14)	&	0.563	\\
%\addlinespace[0.2cm]
$4^+_1$	&	2	&	1.518	&	1.523	&	$4^+_1\rightarrow 2^+_1$	&	$E2$	&	22.6(17)$^{\mathrm{b}}$	&	25.6	\\
		&		&			&			&	$4^+_1\rightarrow 0^+_1$	&	$E4$	&	-			&	1.09	\\
%\addlinespace[0.2cm]
$4^+_2$	&	1	&	2.462	&	2.482	&	$4^+_2\rightarrow 4^+_1$	&	$M1$	&	0.90(18)	&	1.13	\\
		&		&			&			&	$4^+_2\rightarrow 2^+_1$	&	$E2$	&	1.52(19)	&	1.44	\\
		&		&			&			&	$4^+_2\rightarrow 2^+_3$	&	$E2$	&	$<4\cdot 10^3$ 	&	10.5	\\
		&		&			&			&	$4^+_2\rightarrow 0^+_1$	&	$E4$	&	-			&	0.55	\\
%\addlinespace[0.2cm]
$4^+_3$	&	2	&	-		&	2.884	&	$4^+_3\rightarrow 0^+_1$	&	$E4$	&	-			&	0		\\
		&		&			&			&	$4^+_3\rightarrow 4^+_1$	&	$M1$	&	-			&	0		\\
%\addlinespace[0.2cm]
$4^+_4$	&	2	&	-		&	3.025	&	$4^+_4\rightarrow 0^+_1$	&	$E4$	&	-			&	0.84	\\
		&		&			&			&	$4^+_4\rightarrow 4^+_1$	&	$M1$	&	-			&	0		\\

\end{tabular}
\end{ruledtabular}
\end{table}

The experimental and calculated level energies as well as the $M1$ and $E2$ transition strengths are compiled in Table \ref{tab:trans}. As for the level scheme, the transition strengths are in overall agreement. Only the transitions depopulating the $3^+$ states are predicted too strong by about one order of magnitude. For transitions which are forbidden for the applied Hamiltonian, only small transition strengths are observed experimentally. Of particular interest for the investigation of hexadecapole components in the $4^+_{1,2}$ states is the $4^+_2 \rightarrow~4^+_1$ $M1$ transition. The IBM predicts a transition strength of $1.13~\mu_N^2$ which is close to the experimental value of $0.90(18)~\mu_N^2$. No other $4^+$ state is found to show enhanced $M1$ transitions to the $4^+_1$ state in the calculations. Also the $4^+_2\rightarrow~2^+_1$ $E2$ transition strength is reproduced by the model. The predicted $E2$ branching with sizable strength of $10.5~\mathrm{W.u.}$ to the $2^+_3$ state is way below the experimental sensitivity limit.

The predicted $F$-spin quantum numbers are shown in Table \ref{tab:trans} as well. $F$-spin quantum numbers of $F_{\mathrm{max}}-1$ are obtained for the $2^+_3$, $1^+_1$, $3^+_2$, and $4^+_2$ states. From their decay properties, the $2^+_3$ state and the $1^+_1$ and $3^+_2$ states can be identified as the experimentally known one- and two-phonon quadrupole MSS, respectively \cite{Piet01, Klei02}. They will be discussed in an upcoming publication. A mixed-symmetry character is also predicted for the $4^+_2$ state. A variation of the strength parameters $\lambda_{\mathrm{sd}}$ and $\lambda_{\mathrm{sg}}$ revealed, that the $4^+_2$ state is most sensitive to the $\hat{M}_{sg}$ operator. Thus a one-phonon mixed-symmetry hexadecapole character is obtained for the $4^+_2$ state. In contrast, a fully-symmetric character is predicted for the $4^+_1$ state based on the calculated $F$-spin quantum number.

To quantify the amount of $M1$ strength of the $4^+_2\rightarrow~4^+_1$ transition related to the $g$- and $d$-boson parts of the $M1$ operator, the $\left<4^+_1\left||M1\right||4^+_2\right>$ matrix element was recalculated with the values $g_{d_{\pi}}=0$ and $g_{g_{\pi}}=0$, respectively. The respective other value was kept at the value obtained from the parameter scan. With a contribution of $83\mathrm{\%}$ the $\left<4^+_1\left||M1\right||4^+_2\right>$ matrix element is dominated by the $g$-boson part of the $M1$ operator, while only $17\mathrm{\%}$ is related to the $d$-boson part.

\begin{figure}[t!]
	\includegraphics[width=0.48\textwidth]{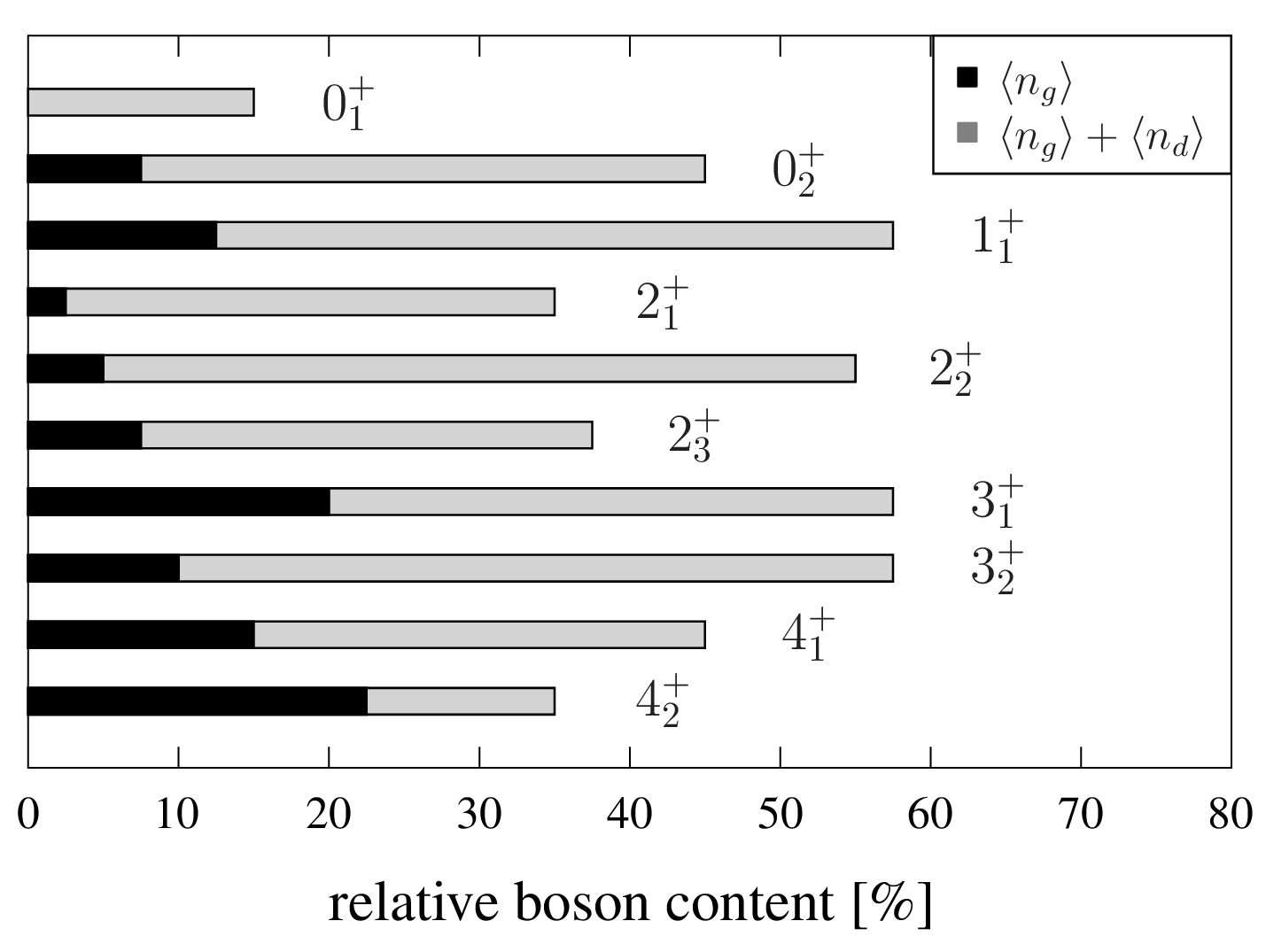}
\caption{\label{fig:bos_config}Calculated $g$- and ($d+g$)-boson contents in the low-lying positive parity states of $^{96}$Ru, indicated with black and gray bars, respectively. The remaining fraction is related to $s$-boson components. Enhanced $g$-boson contents are predicted for the $1^+_1$,  $3^+_1$, $4^+_1$, and $4^+_2$ states.}
\end{figure}

The $d$- and $g$-boson contents of the low-lying, positive-parity low-spin states are shown in Fig. \ref{fig:bos_config}. With a value of $23~\mathrm{\%}$, the largest $g$-boson content is obtained for the $4^+_2$ state, supporting the one-phonon hexadecapole assignment. A similar $g$-boson contribution is predicted for the $3^+_1$ state. This can be explained assuming a dominant $(g^{\dagger}d^{\dagger})^{(3)}$ structure. In this case the $E2$ transition to the $2^+_1$ state would be forbidden by $d$-parity selection rules. This is supported by the calculation and is in remarkable agreement with the data (see Table \ref{tab:trans}). Also for the $4^+_1$ state a large $g$-boson content is obtained which is considerably enhanced compared to, e.g., the $2^+_2$ state, which is known to be the $2^+$ member of the ($2^+_\mathrm{s}\otimes 2^+_\mathrm{s}$) triplet. In addition to the $g$-boson content, a $d$-boson contribution of about $30~\mathrm{\%}$ is predicted by the IBM for the $4^+_1$ state. For the chosen parameter of $\beta$, a mixing of the one-phonon $g$-boson excitation with two-phonon $d$-boson excitations is allowed, which are at similar energies. This is also reflected by the collective $B(E2;4^+_1\rightarrow~2^+_1)$ transition strength. In contrast, the $d$-boson content of the $4^+_2$ state is a factor of 2 less compared to the $4^+_1$ state.

If the enhanced $g$-boson contributions can be attributed to one-phonon hexadecapole contents in the wave functions, this should lead to sizable $E4$ strengths. The $E4$ transition operator was defined in the same way as in \cite{Casp13}, namely

\begin{equation}
\hat{T}(E4)=e_{\pi 1}\left[s_{\pi}^{\dagger}\tilde{g}_{\pi}+g_{\pi}^{\dagger}\tilde{s}_{\pi}\right]^{(4)}.
\end{equation}

Since no $B(E4)$ strengths are known for $^{96}$Ru so far, the $e_{\pi 1}$ value was arbitrarily set to $1~\sqrt{\mathrm{W.u.}}$. With this, $E4$ transition strengths of $1.09$ and $0.55~\mathrm{W.u.}$ are predicted for the $4^+_1$ and $4^+_2$ states, respectively. Their $E4$ strengths are enhanced compared to, e.g., that of the $4^+_3$ state. However, a similar $E4$ transition to the ground state is predicted for the $4^+_4$ state as well. Further constraints for the $E4$ transition operator might be obtained from a measurement of $E4$ strengths, e.g., in $(e,e^{\prime})$ experiments. 

To conclude, the $sdg$-IBM-2 calculations provide strong evidences for MS and FS one-phonon hexadecapole contributions to the lowest-lying $4^+$ states of $^{96}$Ru. However, other mechanisms, such as the $g$ factors of the individual microscopic configurations in their wave functions have to be considered as well as being responsible for the generation of $M1$ strengths between low-lying $4^+$ states. They might be studied within the scope of shell-model calculations with realistic interactions or the quasiparticle phonon model (QPM). 

\begin{figure}[t!]
	\includegraphics[width=0.48\textwidth]{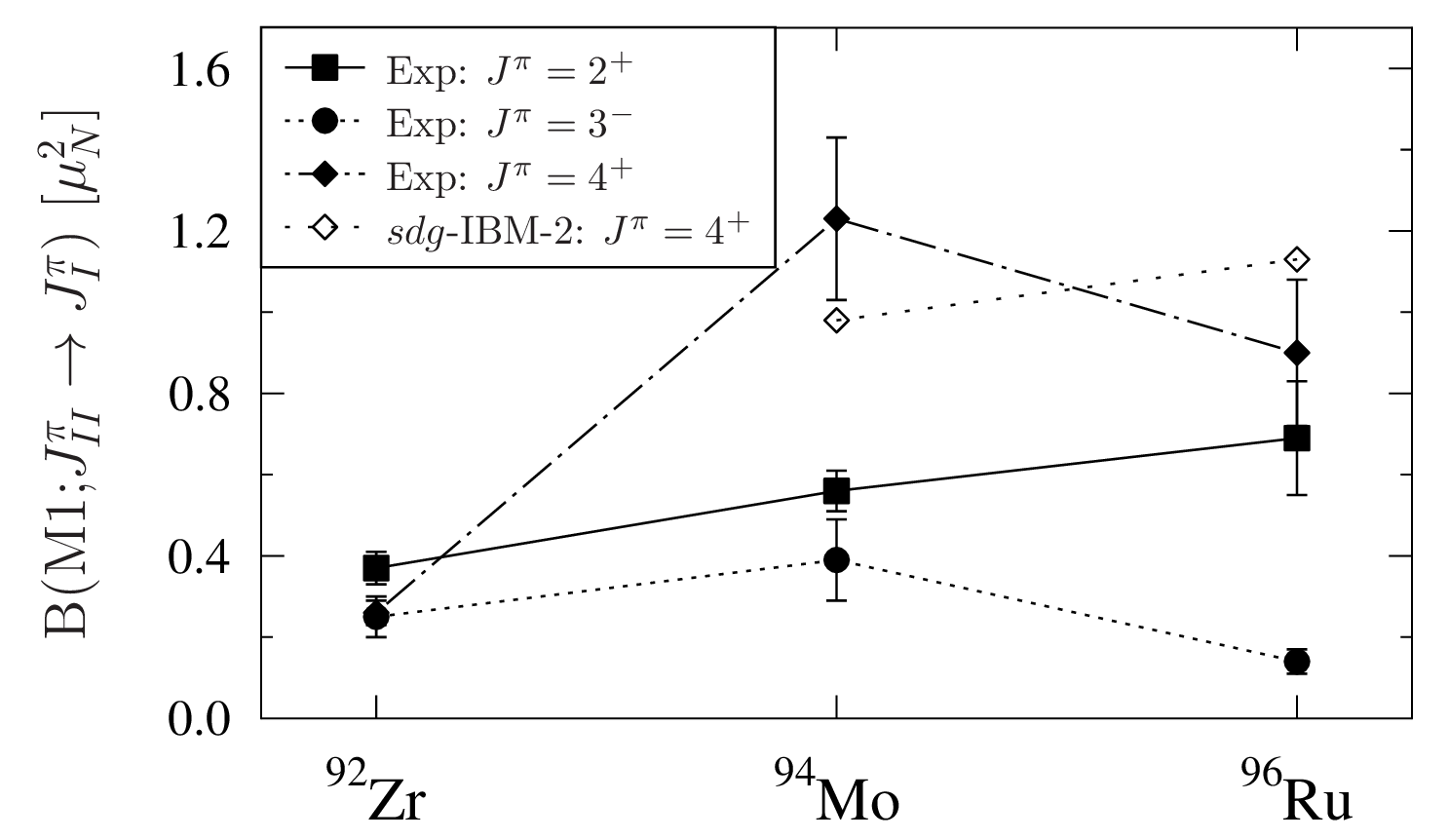}
\caption{\label{fig:m1syst}$M1$ strengths of the one-phonon MSS ($J^{\pi}_{II}$) to FSS ($J^{\pi}_{I}$) transitions in the even-even $N=52$ isotones of quadrupole (full squares), octupole (full circles), and hexadecapole (full diamonds) character. The values obtained in $sdg$-IBM-2 calculations (open triangles) are shown as well. Data for $^{92}$Zr and $^{94}$Mo are taken from \cite{Fran05} and \cite{Fran03}, respectively. The $sdg$-IBM-2 results for $^{94}$Mo are adopted from \cite{Casp13}.}
\end{figure}

\textit{Comparison to $^{92}$Z\MakeLowercase{r} and $^{94}$M\MakeLowercase{o}.} With the new experimental data obtained in this work, one-phonon MSS of quadrupole and possible octupole and hexadecapole character were studied in the $N=52$ isotones as a function of proton number. Figure \ref{fig:m1syst} shows the $M1$ transition strengths of the one-phonon MS to FS states for the different multipolarities for the nuclei $^{92}$Zr, $^{94}$Mo, and $^{96}$Ru. While for the quadrupole states an increase of $M1$ strength is observed with increasing proton number, the $M1$ strength decreases from $^{94}$Mo to $^{96}$Ru for higher multipolarities. It has to be mentioned, that the decrease might be related to a possible fragmentation of the one-phonon octupole and hexadecapole mixed-symmetry states which can not be excluded on the basis of the present experimental data.

The trend for the quadrupole states agrees with shell-model calculations, predicting a maximum $M1$ strength for $^{96}$Ru, based on the concept of configuration isospin polarization (CIP) \cite{Holt07}. Unfortunately, no results on $4^+$ states were reported in Ref. \cite{Holt07}. The decrease of the $M1$ strengths for the $4^+$ states with increasing proton number is not reproduced by the IBM, which predicts a similar trend as for the quadrupole states. 

\textit{Summary.} The observation of strong $M1$ transitions in $^{96}$Ru between the lowest-lying $3^-$ and $4^+$ states provides experimental evidence for one-phonon mixed-symmetry octupole and hexadecapole components in the wavefunctions of the $3^{(-)}_2$ and $4^+_2$ states, respectively. The interpretation on the latter is supported by $sdg$-IBM-2 calculations. Together with the results of Ref. \cite{Casp13}, the new data on $^{96}$Ru suggest that the presence of hexadecapole components in the wave functions of low-lying $4^+$ states is a general phenomenon in near spherical nuclei.

\textit{Acknowledgments.} The authors thank R. Casperson and S. Heinze for support with the IBM calculations and A. Poves for useful discussions. Furthermore, we highly acknowledge the support of the accelerator staff at WNSL, Yale and IKP, Cologne during the beam times. This work is supported by the DFG under grant No.~ZI 510/4-2 and grant No.~SFB 634, the U.S.~Department of Energy grant No.~DE-FG02-01ER40609, and the BMBF grant No.~05P12RDFN8. P.P.~is grateful for the financial support of the Bulgarian Science Fund under contract DFNI-E 01/2. D.R.~and D.S.~acknowledge the German Academic Exchange Service (DAAD) for financial support. S.G.P.~and M.S.~are supported by the Bonn-Cologne Graduate School of Physics and Astronomy.

\bibliography{bibtex/bibtex.bib}

\end{document}